\documentclass[useAMS,usenatbib]{mn2e}
\usepackage{graphicx,xspace,color,longtable}
\usepackage{rotate,amssymb,amsmath,shadow,bezier,curves}

\title[Studying the Dynamical Properties of 20 Nearby Galaxy Clusters]{Studying the Dynamical Properties of 20 Nearby Galaxy Clusters}
\author[Abdullah et al.]{\small{Mohamed H. Abdullah$^1$
, Gamal B. Ali $^1$, H. A. Ismail$^{1,3}$ and Mohamed A. Rassem$^2$}\\
$^1$ Department of Astronomy, National Research Institute of Astronomy and Geophysics, Egypt\\
$^2$ Department of Astronomy, Cairo University, Egypt\\
$^3$ Space and Astronomy department, King Abdul-Aziz University, Jeddah, KSA}

\begin{document}

\maketitle

\begin{abstract}
Using SDSS-DR7, we construct a sample of 42382 galaxies with redshifts in the region of 20 galaxy clusters. 
Using two successive iterative methods, the adaptive kernel method and the spherical infall model, we obtained 
3396 galaxies as members belonging to the studied sample. The 2D projected map for the distribution of the 
clusters members is introduced using the 2D adaptive kernel method to get the clusters centers. The cumulative 
surface number density profile for each cluster is fitted well with the generalized King model. The core radii 
of the clusters' sample are found to vary from 0.18 Mpc $\mbox{h}^{-1}$ (A1459) to 0.47 Mpc $\mbox{h}^{-1}$ (A2670) 
with mean value of 0.295 Mpc $\mbox{h}^{-1}$. 

The infall velocity profile is determined using two different models, Yahil approximation and Praton model. 
Yahil approximation is matched with the distribution of galaxies only in the outskirts (infall regions) of many 
clusters of the sample, while it is not matched with the distribution within the inner core of the clusters. 
Both Yahil approximation and Praton model are matched together in the infall region for about 9 clusters in the 
sample but they are completely unmatched for the clusters characterized by high central density. For these cluster, 
Yahil approximation is not matched with the distribution of galaxies, while Praton model can describe well the 
infall pattern of such clusters.

The integrated velocity dispersion profile shows that there are different behaviors within the cluster's 
virialized region, while it exhibits a flattened out behavior outside the virialized region up to the turnaround 
radius. Under the assumption that the mass follows galaxy distribution, we determine the mass and mass 
profile by two independent mass estimators; projected mass and virial mass methods. The virial mass profile 
is corrected by applying the surface pressure term which reduces the virial mass by about $14\%$. The projected 
mass profile is larger than the corrected virial mass profile for nearly all clusters by about $28\%$. The virial 
mass agrees with NFW mass and Praton mass at $r_v$. The virial mass profile within 1.5 Mpc $\mbox{h}^{-1}$ is fitted 
with NFW mass profile. The concentration parameter ranges from $1.3$ to $39.17$, and has mean value $12.98$. 

\end{abstract}

\section{Introduction}
Knowledge of physics of beginning, evolution, and fate of our universe requires understanding the distribution, 
formation, dynamics, and evolution of matter on a large scale. Galaxy clusters, which are the most massive 
gravitationally bound galaxy systems, play an important role in the study of the large-scale structure formation 
(Fadda et al. 1996; Girardi et al. 1998), as well as to understand the physics of the universe as a whole. 

Studying the properties of galaxy clusters based on old catalogs is affected by the projection effect. 
New and deep redshift surveys (e.g. Sloan Digital Sky Survey, hereafter SDSS) for galaxies on nearly 
whole sky help to overcome such problem. However, the redshift information is distorted by some factors, 
e.g. small scale structure, large scale structure, and observational errors. This distortion leads to difficulty 
to determine the real cluster members which is the most important factor to study the dynamics of galaxy clusters. 
There are many methods used to get cluster members. Some of them are based on statistical rules and others are based 
on the dynamical status of the system. Enhanced methods were introduced in the last 2 decades to get the clusters 
members which take into account the distance from the cluster center beside the redshift information. However, 
no specific method introduces an accurate confirmation on the real cluster members.

The aims of this work are: (i) Determination of centers and members for 20 nearby galaxy clusters based on a 
two-steps method. (ii) Study the redshift space and the infall pattern around them. (iii) Study the 
integrated velocity dispersion profile. (iv) Investigation and determination of the clusters masses and 
mass profiles using two different mass estimators.

The paper is organized as follows. We describe the data sample in Sec. \ref{sec:data}. The determination 
of the cluster center and the selection procedure for cluster membership assignment are described in Sec. 
\ref{sec:cen}. The density profiles of galaxy clusters are illustrated in Sec. \ref{sec:denprof}. The spherical 
infall models; linear model, Yahil approximation and Praton model, are discussed in Sec. \ref{sec:SIM}. The 
velocity dispersion profile and the effects influence it are described on Sec. \ref{sec:veldis}. We briefly 
describe the methods used to compute cluster masses using member galaxies in Sec. \ref{sec:mass}. The clusters 
physical parameters and correlations between them are introduced in Sec. \ref{sec:parameters}. 
We give a brief summary of our main results and draw our conclusions in Sec. \ref{sec:conc}.

\section{Data Sample} \label{sec:data}
We choose 20 nearby Abell clusters from \citealt{Aguerri07} (hereafter AG2007) and Yoon et al. (2008) (hereafter YO2008). 
Using photometric and spectroscopic data of objects classified as galaxies from the SDSS-DR7, we select only those galaxies 
with redshifts between $0.0<z<0.15$ around each galaxy cluster center selected from the two above references. The choice 
of the search radius around the center of each galaxy cluster depends on the value of the cluster redshift taken from 
AG20007 and YO2008. Data for each galaxy consists of right ascension ($\alpha$), declination ($\delta$), redshift, 
and magnitudes u, g, r, i, and z. Table \ref{tab:search} shows the results of the search. Col. 1 is the cluster 
name, col. 2 and 3 are $\alpha$ and $\delta$ of the center of the search, col. 4 is the radius of the search, 
and col. 5 is the number of galaxies found within the search radius. More details about SDSS-DR7 are described 
in  \citealt{Abazajian09} and the website www.sdss.org.

   \begin{table} \centering
   \caption{List of the galaxy clusters sample with the search radius and resultant number.}
   \label{tab:search}
   \begin{tabular}{|c|c|c|c|c|}
   \hline
   Name   	&  $\alpha$ (deg) &	 $\delta$ (deg)	&$\theta$&  No.\\
           	&  (2000)       &	 (2000)	      &    (arcmin)   &      \\ \hline 
   A0117	  &  14.0	      &  -10.0	    &   100	        &  645 \\ 
   A0168	  &  18.7	      &   0.4	      &   180	        &  1296\\
   A0671	  &  127.1	    &   30.4	    &   120	        &  996 \\
   A0779	  &  140.0	    &   33.8	    &   300	        &  4778\\
   A1066	  &  160.0	    &   5.2	      &   100	        &  763 \\
   A1142	  &  165.2	    &   10.6	    &   150	        &  1708\\
   A1205	  &  168.3	    &   2.5	      &   100	        &  890 \\
   A1238	  &  170.7	    &   1.1	      &   100	        &  830 \\
   A1377	  &  176.9	    &   55.8      &   150	        &  2671\\
   A1424	  &  179.4	    &   5.1	      &   100	        &  826 \\
   A1436	  &  180.1	    &   56.2	    &   100	        &  1130\\
   A1459	  &  181.1	    &   1.9       &   300	        &  5960\\
   A1663	  &  195.7	    &   -2.6	    &   100	        &  721 \\
   A1767	  &  204.0	    &   59.2	    &   120	        &  1089\\
   A1809	  &  208.3	    &   5.2	      &   120	        &  1170\\
   A2048	  &  228.8	    &   4.4  	    &   100	        &  1518\\
   A2061	  &  230.3	    &   30.6	    &   100	        &  1365\\
   A2142	  &  239.6	    &   27.2	    &   120	        &  1740\\
   A2255	  &  258.2	    &   64.1	    &   120	        &  813 \\
   A2670	  &  358.6	    &  -10.4	    &   100	        &  502 \\
   \hline\end{tabular}\end{table}   
\section{Determination of the center and members of galaxy cluster} \label{sec:cen}

Dynamical parameters of galaxy clusters, such as mean cluster redshift, velocity dispersion, mass 
and virial and turnaround radii are significantly affected by the method of determination of the 
cluster center and the procedure of the membership selection. The methods used to get cluster 
center and members are illustrated in the following two subsections.

\subsection{Cluster Center}
The cluster center can be defined as, the position at which the surface luminosity is maximum. Depending 
on this definition the cluster center can be determined using either X-ray observation of inter-galactic gas settled 
in the cluster or optical observation of the galaxies themselves. The position of the maximum galaxy number 
per unit area indicates the maximum surface luminosity which by definition indicates the cluster center. 
Also one can define the cluster center as the dynamically oldest part of the cluster so the presence of 
a cD galaxy, giant elliptical galaxy (gE) or a central group of early type galaxies refers to the position 
of the cluster center (den Hartog \& Katgert 1996).

In order to determine the cluster center we apply a non-parametric density estimator called the adaptive 
kernel method, hereafter AKM, (see Pisani 1993, Pisani 1996, Fadda et al. 1998) to get separately the maximum 
probability in $\alpha$, $\delta$, and $z$ directions. Then, the galaxy closest to the maximum probability 
is considered as the cluster center and referred to as the reference galaxy.

\subsection{Membership Determination}

  \begin{figure*}
  \includegraphics[width=20cm,height=11cm]{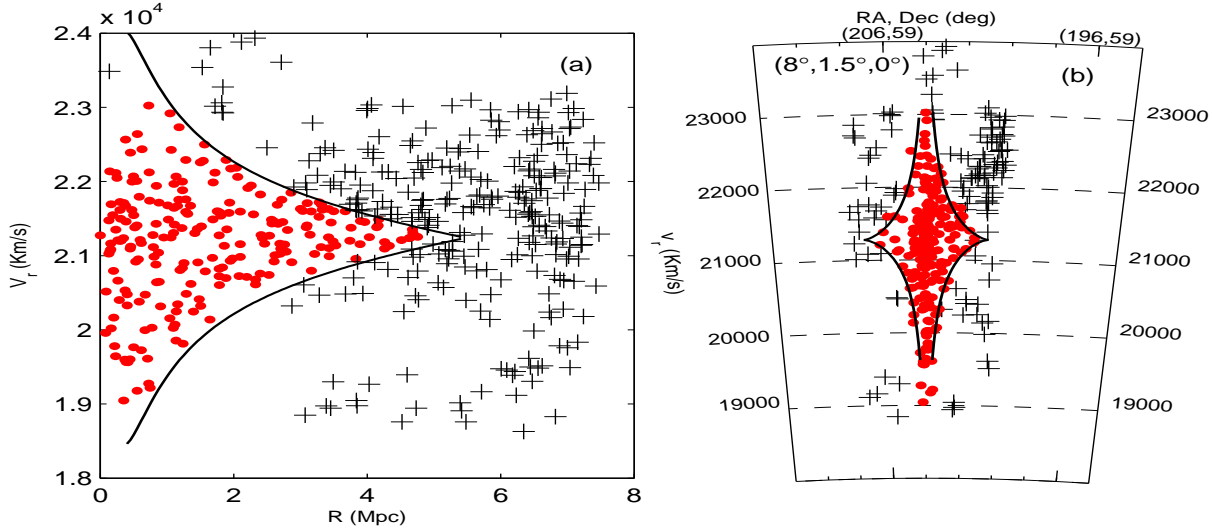} \vspace{-3 cm}
  \caption{The distribution of galaxies of A1767 in $\mbox{S}_{p}$ (panel a) and $\mbox{S}_{v}$ (panel b), respectively. The two curves in each space    indicate the application of SIM.} 
  \label{fig:S1S2}
  \end{figure*}

Several methods have been developed in order to obtain reliable members of galaxy cluster and to avoid the 
presence of interlopers. These methods can be classified into two families. First, those algorithms that 
use only the redshift information, e.g. 3$\sigma$-clipping techniques (Yahil \& Vidal 1977), fixed gapping 
procedures (\citealt{Beers90}, Zabludoff et al. 1990), and jackknife technique (Perea et al. 1990). These 
methods are based on statistical rules. The other family uses information of both position and redshift, 
such as the shifting gap procedure designed by Fadda et al. (1996) or methods designed by den Hartog \& 
Katgert (1996), or Regos \& Geller (1989), which are based on physical rules.

To determine cluster members we use a two-steps method which is described as the following:

1. Redshift Information: AKM is used to find the significant peaks in the redshift distribution. For each iteration 
we choose a cut-off range that is selected manually outside the main peak to secure the conservation of the 
galaxy cluster members. The iteration is stopped when one single-peak is observed within the new cut-off range. 

2. Redshift-Distance Information: Galaxies belonging to the main peak are analyzed in the second step, 
in which we use the combination of the projected distance from the cluster center and redshift information. 
The method used for this purpose is the spherical infall model, hereafter SIM (see Sec. \ref{sec:SIM}). 
After applying SIM one can get the limits of the infall velocity that galaxies would have 
within the cluster as a function of distance from its center. Any galaxy outside this limits 
is classified as an interloper. This method is iterated until no more galaxies are excluded. 
The important advantages of this method are it takes combined radial velocity-space information 
into account and the removing of outliers is based on physical rules. The disadvantage of this 
method is  that it is not valid inside or near the virialized region (see e.g. Regos \& Geller 1989).

Using the redshift space (peculiar velocity versus projected distance from the cluster center, hereafter 
$\mbox{S}_p$), van Haarlem  et al. (1993) conclude that Coma cluster is too elongated to be described by SIM. 
Also, Diaferio \& Geller (1997) say that SIM is not good to describe the infall region 
of galaxy clusters due to random motions (substructures and/or recent mergers). Figure (\ref{fig:S1S2}.a) 
shows $\mbox{S}_p$ for A1767 which describes this problem. As shown, although there are some galaxies (pluses) 
that should be included to the cluster according to Diaferio technique (see Figure 5 in Rines \& Diaferio 2006 
for A1767), SIM (the two curves) does not include them. Another redshift space, called cone diagram or slice 
oriented around the cluster center ($\alpha = 204^{\circ}$ and $\delta = 59^{\circ}$) with length = $8^{\circ}$ 
and width = $1.5 ^{\circ}$ (see Praton \& Schneider 1994), hereafter $\mbox{S}_{v}$, is shown in Figure 
(\ref{fig:S1S2}.b). This figure shows that these galaxies (pluses) which seem to belong to the cluster in 
$\mbox{S}_p$ are far from the cluster in $\mbox{S}_{v}$ and this is due to the projection effect. 
The application of SIM in $\mbox{S}_{v}$ (see Praton 1993) shows that the galaxies which are considered to be 
outliers in $\mbox{S}_p$ are also far and not included within the model in $\mbox{S}_{v}$. Because of that 
and with avoiding the effect of substructure for the studied clusters (see Figure \ref{fig:z2}), we depend on 
SIM to get the cluster members in $\mbox{S}_p$.

As illustrated, we determine clusters' members using the two-steps method, AKM (first step) and SIM (second step). 
We depend on Yahil approximation (see Sec. \ref{sec:SIM}) which requires determination of the density contrast profile, 
$\Delta(\leq R)$, of the cluster, the background density, $\rho_{bg}$, and the cosmological parameter, $\Omega_0$. 
We select $\Omega_0=1$ and $\mbox{H}_0=100 \mbox{ Km s}^{-1} \mbox{ Mpc}^{-1}$. The distance to a cluster center 
is calculated using $D=c z_{cl}/H_0$, where $z_{cl}$ is the average redshift of the cluster's members. Table \ref{tab:z1} 
shows the results of determination of the cluster members after applying these two steps, respectively.

\begin{table}\centering 
\caption{The number of the cluster members after applying the one dimensional AKM (the first step) and SIM (the second step).}
\label{tab:z1}
\begin{tabular}{|c|ccc|c|ccc|} \hline
&\multicolumn{3}{c|}{First Step}&&\multicolumn{3}{c|}{Second Step}\\ 
&\multicolumn{3}{c|}{(AKM)}&&\multicolumn{3}{c|}{(SIM)}\\
\cline{2-4} \cline{6-8}\\
Name&$z_{min}$&$z_{max}$&No.&&$z_{min}$&$z_{max}$&No.\\ \hline
A0117 & 0.047 &	0.060  & 221 && 0.049 &0.059	& 146\\ 
A0168 & 0.037 &	0.053  & 322 && 0.039 &0.050	& 194\\ 
A0671 & 0.045 &	0.057  & 216 && 0.046 &0.057	& 162\\ 
A0779 & 0.018 & 0.031  & 500 && 0.019 &0.031	& 191\\ 
A1066 & 0.061 &	0.079  & 291 && 0.063 &0.075	& 160\\ 
A1142 & 0.028 & 0.040  & 223 && 0.031 &0.039	& 122\\ 
A1205 & 0.067 &	0.089  & 316 && 0.072 &0.080	& 93 \\ 
A1238 & 0.065 &	0.078  & 190 && 0.070 &0.078	& 85 \\ 
A1377 & 0.048 &	0.055  & 353 && 0.048 &0.055	& 158\\ 
A1424 & 0.069 &	0.082  & 313 && 0.071 &0.080	& 94 \\ 
A1436 & 0.058 &	0.071  & 278 && 0.059 &0.070	& 142\\ 
A1459 & 0.015 &	0.025  & 284 && 0.016 &0.025	& 189\\ 
A1636 & 0.077 &	0.094  & 428 && 0.079 &0.089	& 110\\ 
A1767 & 0.062 &	0.081  & 511 && 0.064 &0.077	& 226\\ 
A1809 & 0.074 &	0.087  & 418 && 0.075 &0.085	& 125\\ 
A2048 & 0.086 &	0.109  & 401 && 0.092 &0.103	& 90 \\ 
A2061 & 0.067 &	0.089  & 486 && 0.071 &0.083	& 228\\ 
A2142 & 0.084 &	0.100  & 746 && 0.084 &0.098	& 419\\ 
A2255 & 0.072 &	0.087  & 439 && 0.073 &0.087	& 279\\ 
A2670 & 0.068 &	0.085  & 309 && 0.070 &0.082	& 183\\ 
\hline 
\end{tabular} \end{table} 

  \begin{figure*}
  \includegraphics[width=22cm,height=24cm]{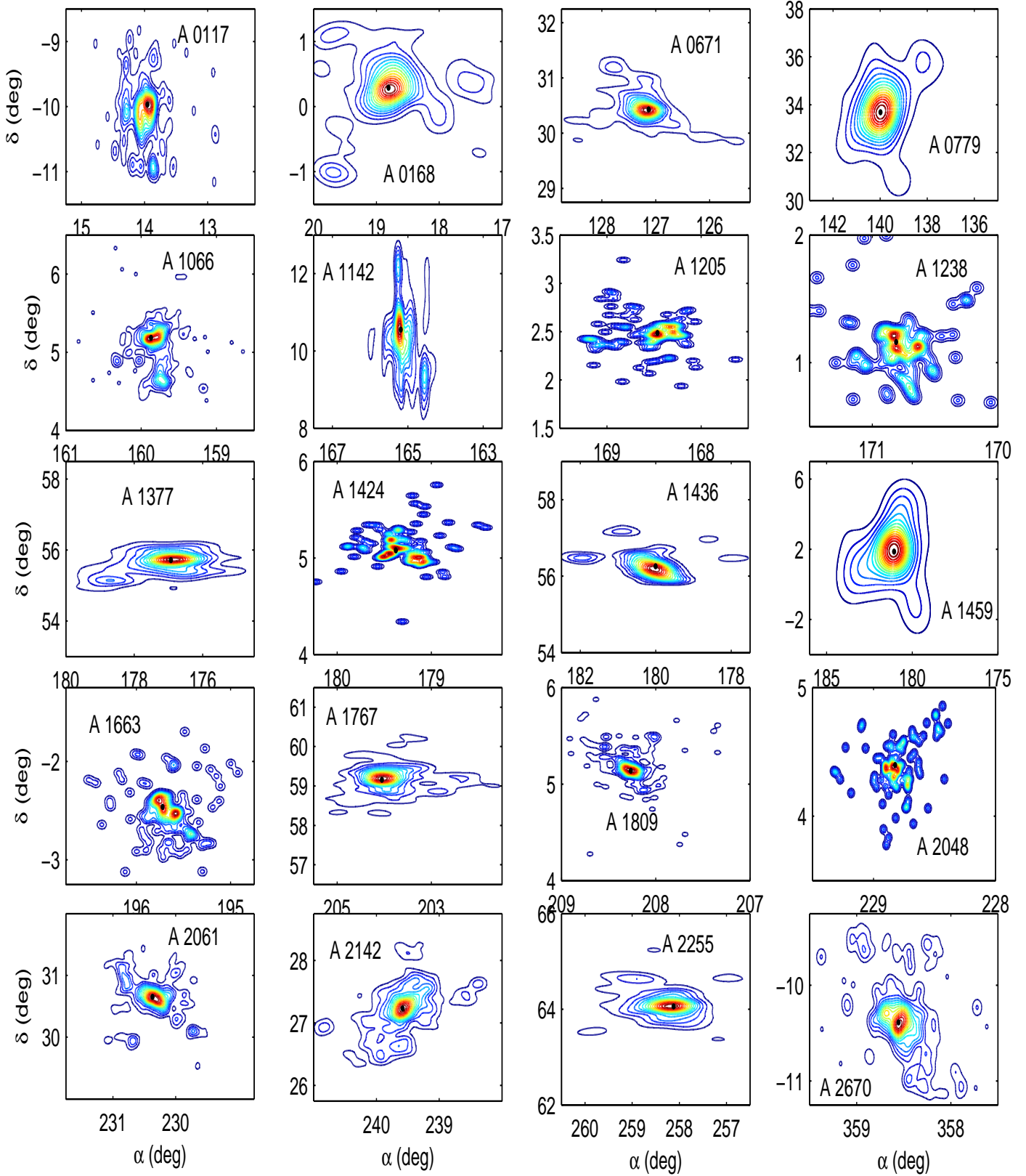} \vspace{-2cm}
  \caption{Adaptive-kernel contour map (20 levels) of surface number density. Filled circle represents the galaxy cluster center.} 
  \label{fig:con}
  \end{figure*}
 
The center of each cluster is obtained using the two dimensional AKM applied to $\alpha$ and $\delta$ 
beside the one dimensional AKM applied to $z$ for the galaxies considered as cluster members. Figure 
\ref{fig:con} shows the isodensity contour maps for the distribution of the cluster members. The four 
clusters A0168, A0671, A0779 and A1459 exhibit smooth distributions of galaxies without substructure in 
the projected map, while the other clusters show non-regularity in the isodensity map distribution and 
some of them show presence of substructures in their projected maps. 

Figure \ref{fig:z2} shows the redshift distribution for galaxy members in each cluster. We find that most 
clusters appear as a well isolated peak in the redshift space and represent Gaussian distribution. This means 
that there are no substructures in the redshift space within these clusters, although in the projected map 
some of them show presence of substructure.

  \begin{figure*}
  \includegraphics[width=19cm,height=18cm]{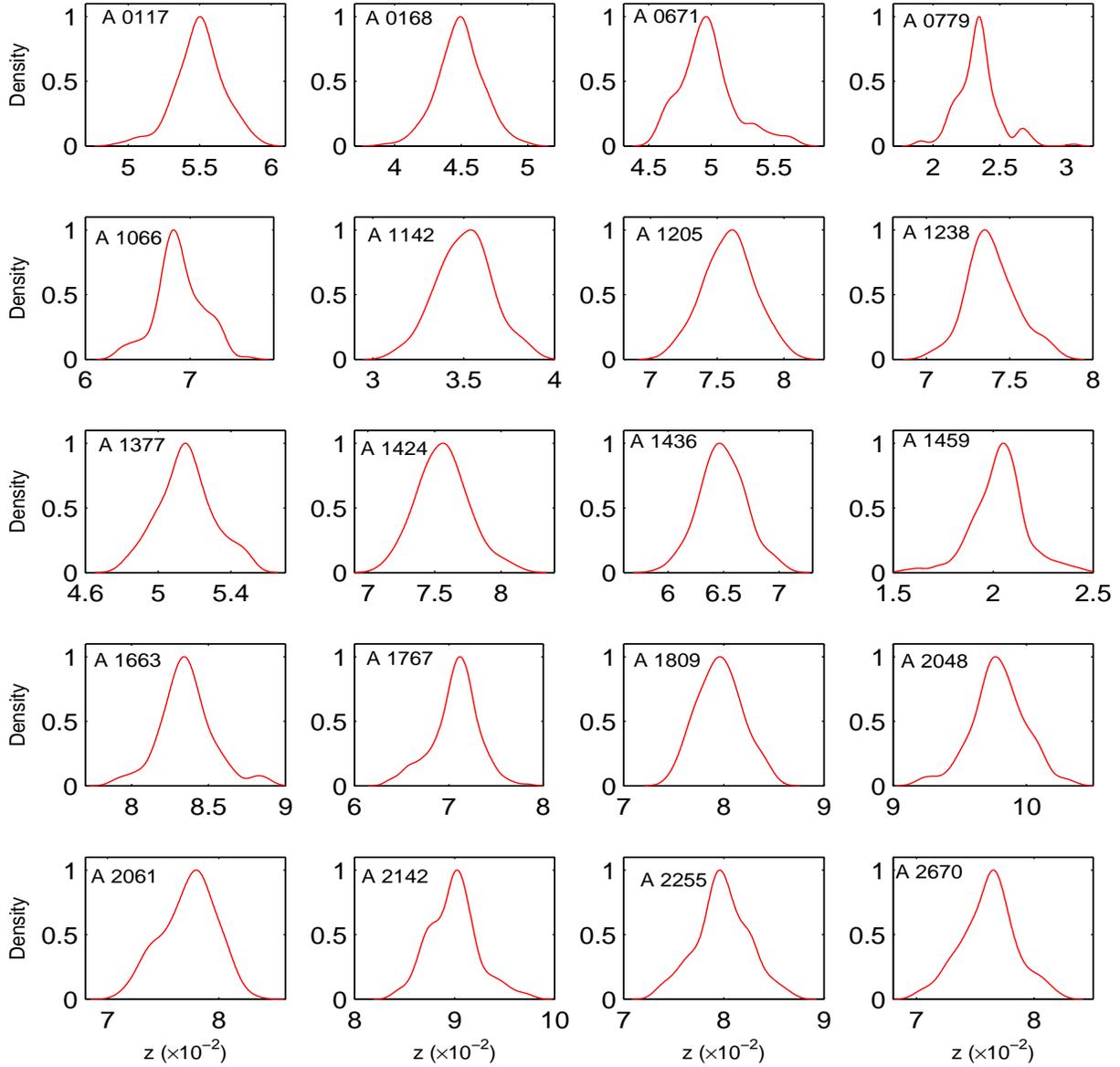} \vspace{-1.5cm}
  \caption{Redshift distribution of galaxies in each cluster of the studied sample.} 
  \label{fig:z2}
  \end{figure*}

\begin{figure}
  \includegraphics[width=11cm,height=10cm]{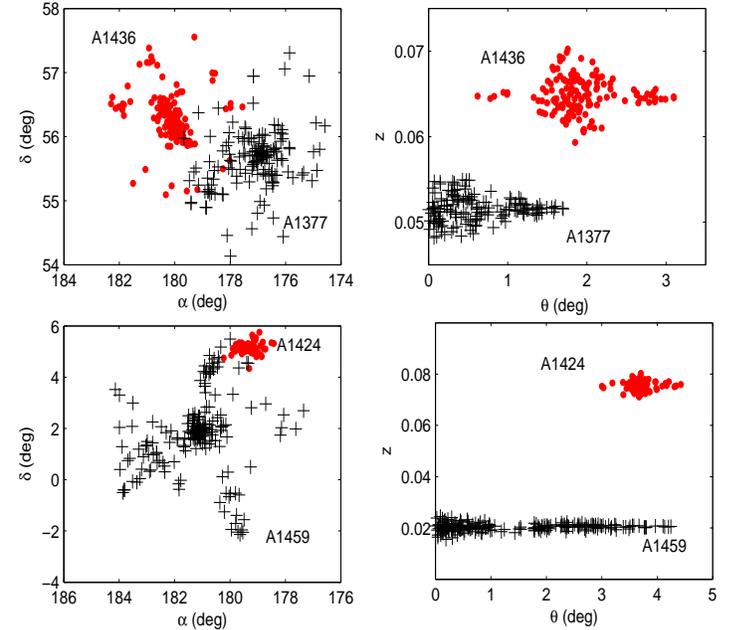}\vspace{-1cm}
  \caption{Sky position (left panels) and redshift space (right panels) for the clusters appearing overlapped: 
  clusters A1377 (asterisk) and A1436 (point) (upper panels) and clusters A1424 (point) and 1459 (asterisk) (lower panels).}
  \label{fig:doub}
  \end{figure}

It is found that there are projection overlaps between the two clusters A1377 and A1436 and the two clusters A1424 and A1459. 
To avoid duplication of any member among such overlapped clusters, we explore them in redshift space, see Figure \ref{fig:doub}. 
The two clusters A1377 and A1436 are plotted in the redshift space relative to A1377 center and the redshift space of the two 
clusters A1424 and A1459 is drawn relative to A1459 center. Although the sky map (Figure \ref{fig:doub} left panels) of the 
members of each cluster pairs shows overlapping, the redshift space map (Figure \ref{fig:doub} right panels) shows that the
clusters in each pair are separate.

Table \ref{tab:centers} shows the basic parameters of each cluster. Cols. 2-4 give the cluster center using 
AKM, cols. 5-7 give the coordinates of the reference galaxy, col. 8 gives the mean redshift of the cluster, 
and cols. 9-11, and 12-14 give $\alpha$, $\delta$, and $<z>$ for AG2007 and YO2008, respectively (see also 
Rines et al. 2006). It is clear that the clusters centers obtained by this study are very close to those work.

\begin{table*} 
\caption{The basic parameters of the clusters in the studied sample with comparison with other studies.}
\label{tab:centers}
\begin{tabular}{|c|ccc|c|ccc|c|ccc|cccc|} \hline
&\multicolumn{3}{c||}{AKM}&&\multicolumn{3}{c||}{Ref. Galaxy}&&\multicolumn{3}{c||}{AG2007}&&\multicolumn{3}{c||}{YO2008}\\
\cline{2-4}\cline{6-8}\cline{10-12}\cline{14-16}\\ 
Name&$\alpha$(deg)&$\delta$(deg)&z&&$\alpha$ (deg)&$\delta$ (deg)&z&$<z>$&$\alpha$ (deg)&$\delta$ (deg)&$<z>$&&$\alpha$ (deg)&$\delta$ (deg)&$<z>$\\
\hline 
A0117&13.96	  &-10.02 & 0.055 &&13.96 &-9.97	& 0.055&0.055& 14.01 &-10.00&0.055&&14.08&-10.00 &0.055\\ 
A0168&18.79	  &0.28	  & 0.045 &&18.81 & 0.29	& 0.046&0.045& 18.74 &0.37  &0.045&&-     &-     &-    \\ 
A0671&127.17	&30.41	& 0.050 &&127.13& 30.43& 0.050&0.050& 127.12&30.42 &0.049&&127.13&30.41 &0.051\\ 
A0779&139.95	&33.68	& 0.023 &&139.99& 33.68& 0.023&0.023& 139.96&33.77 &0.023&&-     &-     &-    \\ 
A1066&159.77	&5.19	  & 0.068 &&159.87& 5.18	& 0.068&0.069& 159.91&5.174 &0.069&&159.77&5.21  &0.069\\ 
A1142&165.21	&10.44	& 0.035 &&165.19& 10.55& 0.034&0.035& 165.22&10.55 &0.035&&-     &-     &-    \\ 
A1205&168.41	&2.47	  & 0.076 &&168.46& 2.49	& 0.073&0.076& 168.32&2.54  &0.076&&168.33&2.54  &0.076\\ 
A1238&170.77	&1.11	  & 0.073 &&170.81& 1.16	& 0.073&0.074& 170.71&1.09  &0.074&&-     &-     &-    \\ 
A1377&176.88	&55.72	& 0.051 &&176.93& 55.71& 0.051&0.052& 176.88&55.76 &0.051&&176.81&55.72 &0.052\\ 
A1424&179.34	&5.06	  & 0.076 &&179.38& 5.08	& 0.076&0.075& 179.36&5.12  &0.076&&179.37&5.09  &0.076\\ 
A1436&180.03	&56.21	& 0.065 &&180.00& 56.26& 0.067&0.065& 180.09&56.23 &0.065&&180.06&56.26 &0.064\\ 
A1459&181.06	&1.86	  & 0.021 &&181.06& 1.87	& 0.021&0.020& 181.10&1.88  &0.020&&-     &-     &-    \\ 
A1663&195.69	&-2.49	& 0.083 &&195.72& -2.46& 0.080&0.084& 195.71&-2.52 &0.083&&195.72&-2.41 &0.084\\ 
A1767&204.03	&59.21	& 0.071 &&204.05& 59.16& 0.071&0.071& 204.02&59.20 &0.071&&204.03&59.21 &0.071\\ 
A1809&208.26	&5.15	  & 0.080 &&208.27& 5.14	& 0.079&0.080& 208.24&5.16  &0.079&&208.26&5.14  &0.080\\ 
A2048&228.83	&4.34	  & 0.098 &&228.82& 4.40	& 0.098&0.098& -     &-     &-    &&228.81&4.38  &0.098\\ 
A2061&230.34	&30.65	& 0.078 &&230.37& 30.65& 0.079&0.077& 230.31&30.61 &0.079&&230.33&30.67 &0.079\\ 
A2142&239.58	&27.26	& 0.090 &&239.58& 27.23& 0.091&0.090& -     &-     &-    &&239.58&27.23 &0.090\\ 
A2255&258.17	&64.07	& 0.080 &&258.12& 64.06& 0.073&0.080& 258.22&64.07 &0.080&&258.20&64.05 &0.083\\
A2670&358.52	&-10.36	& 0.077 &&358.56&-10.39& 0.079&0.076& 358.55&-10.41&0.076&&358.55&-10.39&0.076\\ 
\hline 
\end{tabular} \end{table*}
\section{Density Profiles of Galaxy Clusters} \label{sec:denprof}

The number density profile of galaxy clusters is introduced by some authors to describe the surface 
distribution of galaxies in clusters. King (1966, 1972) introduced an analytical representation of 
the galaxy distribution profile in clusters. This profile was devised to describe the surface 
brightness distributions in globular clusters and in elliptical galaxies, but it can also be 
applied to galaxy clusters (\citealt{Bahcall77}). The generalized King profile is given by

   \begin{equation} \label{eq:king1}
   \Sigma(<R)= \Sigma_0\left(1+\frac{R^2}{r_c^2}\right)^{\gamma}+\Sigma_{bg}
   \end{equation}

\noindent where $\Sigma_0$ is the central number density per unit area, $r_c$ is the core radius, 
$\gamma$ is a power parameter and $\Sigma_{bg}$ is the background surface number density of the universe (\citealt{Adami98}). 
The corresponding spatial density profile, using Abel integral, is given by

   \begin{equation} \label{eq:king2}
   \rho(<r)=\rho_0\left(1+\frac{r^2}{r_c^2}\right)^{(\gamma-0.5)}+ \rho_{bg},
   \end{equation}   

   \begin{equation} \label{eq:king3}   
   \rho_0=\frac{\Gamma(0.5-\gamma)}{\Gamma(0.5)\Gamma(-\gamma)}\frac{\Sigma_o}{r_c}
   \end{equation}

\noindent (Regos \& Geller 1989 and Van Haarlem et al. 1993). Note that $\rho_0$ is the central 
number density per unit volume and $\rho_{bg}$ is the background volume number density of the universe.

The background density, $\rho_{bg}$, can be obtained by integrating the Schechter luminosity function 
(Schechter 1976) and its corresponding magnitude function, $\Phi(M)$, where

   \begin{equation} \begin{split}
   \Phi(M)dM& = 0.921N^*\left(10^{0.4(\alpha+1)(M^*-M)}\right)\times\\
            & \quad \exp\left(-10^{0.4(M^*-M)}\right)dM,
   \end{split} 
   \end{equation}

   \begin{equation}\label{eq:lum4} \begin{split}
   \rho_{bg} &     = \int_{-\infty}^{M_{lim}} \Phi(M)dM \\           
            &\quad = 0.921N^* \Gamma \left(\alpha+1,10^{0.4(M^*-M_{lim}+5\log D+25)}\right)
   \end{split}
   \end{equation}    

\noindent where $M_{lim}$ is the limiting apparent magnitude of the survey and $D$ is the distance to the 
cluster center in Mpc (Regos \& Geller 1989). The three parameters $N^*$, $M^*$ and $\alpha$ are determined 
for the the survey using Schechter luminosity function. We choose $N^*=0.0468 \mbox{ h}^3\mbox{ Mpc}^{-3}$, 
$M^*=-18.24+5\log_{10}\mbox{h}$, $\alpha= -1.31$ for $u$ band with cosmological parameters $\Omega_m=1.0$ 
and $\Omega_{\Lambda}=0.0$ (\citealt{Blanton01}).

The mass density profiles of galaxy clusters are introduced by several authors. Navarro et al. 
(1995, 1996, 1997, hereafter NFW), and \citealt{Hernquist90} propose two-parameter models based on Cold 
Dark Matter (CDM) simulations of haloes. The mass density profile introduced by NFW is given by

  \begin{equation} \label{eq:NFW1}
  \rho(r)=\frac{\delta_s \rho_s}{\frac{r}{r_s}\left(1+\frac{r}{r_s}\right)^2}, \end{equation}

\noindent and its corresponding mass profile is given by

   \begin{equation} \label{eq:NFW2}
   M(<r)=\frac{M_s}{\ln(2)-(1/2)}\left[ln(1+\frac{r}{r_s})-\frac{r/r_s}{1+r/r_s}\right]
   \end{equation}

\noindent where $r_s$ is the scale radius, 
$M_s=4\pi\delta_s\rho_s r^3_s [\ln(2)-(1/2)]$ is the mass within $r_s$, and $\delta_s$ is 
the characteristic density (see Koranyi \& Geller 2000, Rines et al. 2003).

The observed cumulative surface number density is fitted with the generalized King model (Eq. \ref{eq:king1}) 
using the Curve Fitting MatLab Toolbox. The parameters obtained from the fit are listed in Table \ref{tab:Kingfitting}. 
Col. 2 gives the cluster core radius, $r_c$, col. 3 gives the central surface number density, $\Sigma_0$, and col. 4 
gives the value of $\gamma$. The standard error for each fitted parameter is with $95\%$ confidence level. 
Col. 5 gives the adjusted R-square to indicate the goodness of fit (see MatLab help for the goodness of fit). 
The adjusted R-square statistic can take on any value less than or equal to 1, with a value closer to 1 
indicating a better fit.The core radius, $r_c$, has mean value of 0.295 Mpc $\mbox{h}^{-1}$ and range from 
0.18 Mpc $\mbox{h}^{-1}$ (A1459) to 0.47 Mpc $\mbox{h}^{-1}$ (A2670). While $\Sigma_0$ has mean value 146.29 
Mpc$^{-2}$ $\mbox{h}^{2}$ and range from 53.6 Mpc$^{-2}$ $\mbox{h}^{2}$ (A1424) to 411.8 Mpc$^{-2}$ $\mbox{h}^{2}$ 
(A1459). Notice that the two clusters A0779 and A1459 are characterized by high central surface densities.

\begin{figure*}
\includegraphics[width=20cm,height=19cm]{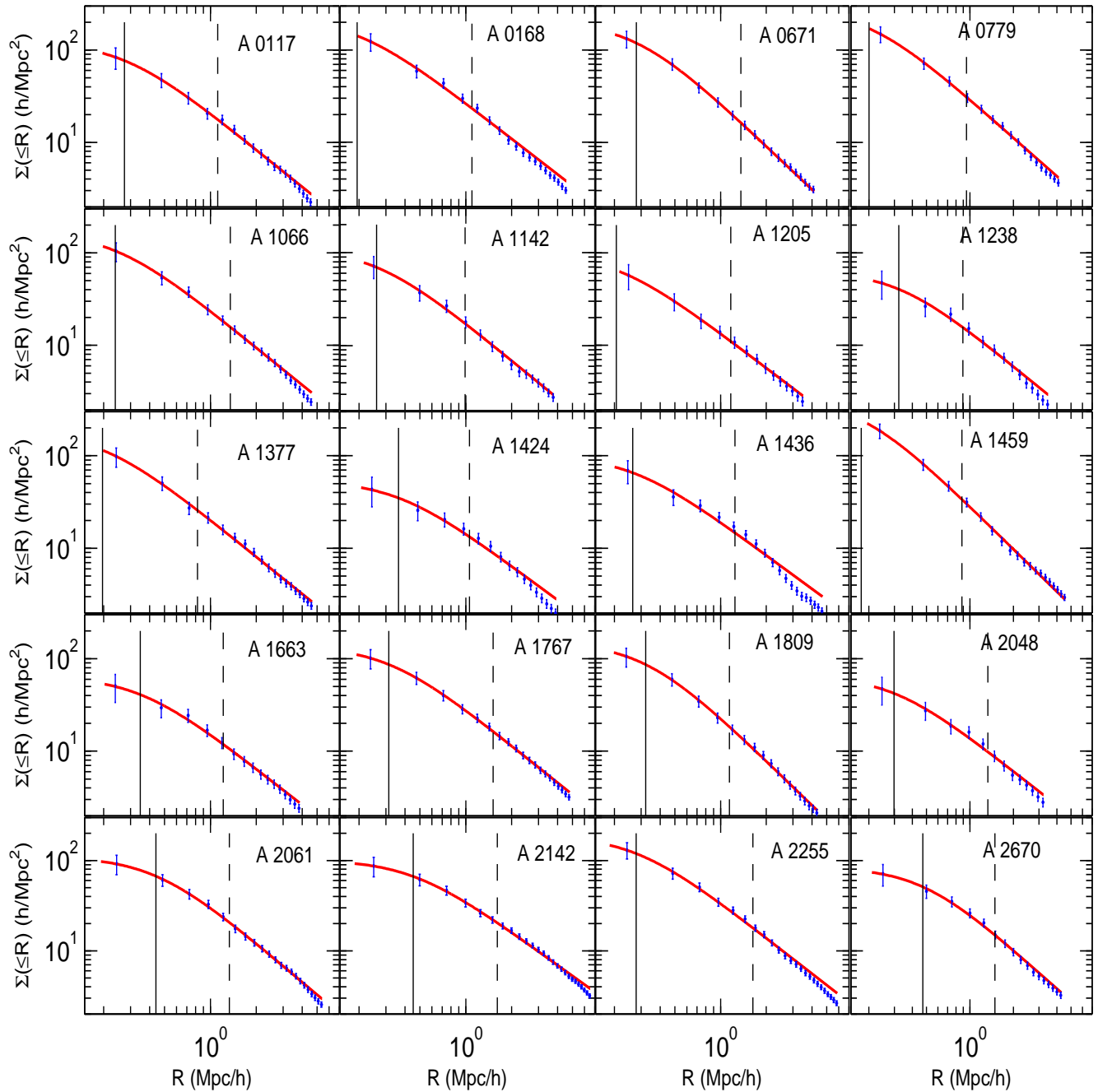} \vspace{-.5cm}
\caption{The cumulative surface number density profiles (dots) together with the generalized King model (solid line). 
The two vertical solid and dashed lines are $r_c$ and $r_v$, respectively.}
\label{fig:K}
\end{figure*}

Figure \ref{fig:K} shows the cumulative surface number density profile (dotted curve) fitted with the 
generalized King model (solid line). The two vertical solid and dashed lines represent the core and virial 
radii (see Sec. \ref{sec:SIM}), respectively. The density decreases rapidly within the central regions 
and then decreases slowly until reaching the boundary of the cluster.

\begin{table} \centering
\caption{Fitted parameters for the King model.}
\label{tab:Kingfitting}
\begin{tabular}{|c|c|c|c|c|}
\hline
Name  &$r_c$            &$\Sigma_0$       &$\gamma$       &R-Square   \\
      &(Mpc/h)          &(h$^2$/Mpc$^{2}$)&               &           \\\hline
A0117	&0.27$\pm$0.03	&122 $\pm$ 6  &-0.68 $\pm$0.03  &0.999 $\pm $ 0    \\
A0168	&0.19$\pm$0.07	&223 $\pm$ 49 &-0.65 $\pm$0.06  &0.996 $\pm $ 0.001\\
A0671	&0.28$\pm$0.02	&205 $\pm$ 8  &-0.79 $\pm$0.03  &0.999 $\pm $ 0    \\
A0779	&0.20$\pm$0.03	&276 $\pm$ 24 &-0.70 $\pm$0.03  &0.999 $\pm $  0   \\
A1066	&0.24$\pm$0.04	&173 $\pm$ 17 &-0.69 $\pm$0.04  &0.998 $\pm $ 0.001\\
A1142	&0.26$\pm$0.06 	&111 $\pm$ 14 &-0.68 $\pm$0.07  &0.997 $\pm $ 0.001\\
A1205	&0.21$\pm$0.03	&100 $\pm$ 10 &-0.63 $\pm$0.03  &0.999 $\pm $ 0    \\
A1238	&0.32$\pm$0.12	&63  $\pm$ 10 &-0.65 $\pm$0.12  &0.991 $\pm $ 0.004\\
A1377	&0.20$\pm$0.04	&182 $\pm$ 26 &-0.67 $\pm$0.04  &0.998 $\pm $ 0.001\\
A1424	&0.36$\pm$0.12	&54  $\pm$ 7  &-0.62 $\pm$0.11  &0.988 $\pm $ 0.005\\
A1436	&0.26$\pm$0.11	&100 $\pm$ 21 &-0.61 $\pm$0.10  &0.999 $\pm $ 0    \\
A1459	&0.18$\pm$0.02	&412 $\pm$ 35 &-0.77 $\pm$0.02  &0.999 $\pm $ 0    \\
A1663	&0.35$\pm$0.10	&64  $\pm$ 7  &-0.66 $\pm$0.10  &0.992 $\pm $ 0.003\\
A1767	&0.31$\pm$0.02	&137 $\pm$ 3  &-0.67 $\pm$0.02  &0.999 $\pm $ 0    \\
A1809	&0.32$\pm$0.02	&151 $\pm$ 5  &-0.81 $\pm$0.03  &0.999 $\pm $ 0    \\
A2048	&0.30$\pm$0.10	&65  $\pm$ 10 &-0.62 $\pm$0.09  &0.994 $\pm $ 0.002\\
A2061	&0.44$\pm$0.03	&111 $\pm$ 3  &-0.73 $\pm$0.03  &0.999 $\pm $ 0    \\
A2142	&0.45$\pm$0.03	&102 $\pm$ 3  &-0.62 $\pm$0.02  &0.999 $\pm $ 0    \\
A2255	&0.28$\pm$0.04	&189 $\pm$ 12 &-0.66 $\pm$0.04  &0.998 $\pm $ 0.001\\
A2670	&0.47$\pm$0.08	&84  $\pm$ 5  &-0.73 $\pm$0.08  &0.996 $\pm $ 0.001\\
\hline 
\end{tabular} \end{table} 

\section{Spherical Infall Model} \label{sec:SIM}

Before describing SIM we describe galaxy clusters themselves. Galaxy 
clusters can be divided into two regions: inner virialized region and outer infall one. A virialized 
region is the region in which the system is in dynamical equilibrium, and the limit of this 
region is the virial radius, $r_v$. This radius can be defined as, the radius within which 
the density is 200 times the critical density of the universe (\citealt{Carlberg97}). 
The virialized region is surrounded by infall region in which the galaxies are falling into the 
gravitational potential well of the cluster, and they have not yet reached equilibrium (Rines et al. 2003).
 
SIM has been extensively described in literature (Gunn \& Gott 1972, 
Silk 1974, Peebles 1976, Schechter 1980, Vedel \& Hartwick 1998). It assumes that galaxy clusters 
started as small density perturbations in the early universe. These perturbations eventually deviate 
from the general Hubble flow of the universe and after reaching a maximum radius, i.e. its turnaround 
radius, they start collapsing. SIM describes the dynamics of the non-equilibrium 
region of galaxy clusters where the effects of virialization and crossing shells are negligible. 
Under the spherical symmetry assumption, the infall motion produces a pattern of caustic shape in 
the galaxy cluster redshift space. This pattern envelops all galaxies whose infall motion overwhelms 
the Hubble flow (Kaiser 1987).

Under the assumption of spherical symmetry, the cluster mass distribution can be considered as a 
set of concentric mass shells, whose centers coincide with that of the density perturbation and 
only the mass inside a shell influences the evolution of that shell, i.e. no mass transformation 
between shells. Hence each mass shell can be considered as an independently evolving Friedmann 
universe, characterized by the density inside it (den Hartog \& Katgert 1996). Consequently, any 
shell enclosing density greater than the critical density will expand to a certain radius and then 
infall toward the cluster. In the frame of the cluster the infall velocity can be defined as

\begin{equation}  
v_{inf}(r)=H_0r+v_{pec}(r)
\end{equation}

\noindent where $v_{pec}$ is the peculiar velocity (see Sec. \ref{sec:veldis}), and $\mbox{H}_0$ 
is the present value of the Hubble parameter (Regos \& Geller 1989). 
Accordingly, the turnaround radius, $r_t$, for each shell can be defined as the radius at 
which the peculiar velocity of that shell exactly cancels the Hubble expansion, so that the 
material at this shell is physically standing still with respect to the cluster (Praton \& Schneider 1994)

The linear theory of density perturbations shows that a spherically symmetric mass concentration 
in an expanding universe induces a radial peculiar velocity field in the surrounding region (see Gunn 1978, Peebles 1980)

\begin{equation}\label{eq:linear}
\frac{v_{pec}}{H_0r}=-\frac{1}{3}\Omega_0^{0.6}\Delta(<r)
\end{equation}

\noindent where $\Delta(<r)$ is the density contrast within radius $r$ which is defined as

\begin{equation}
\Delta(<r)=\frac{\rho(<r)}{\rho_{bg}}-1=\frac{\rho(<r)}{\Omega_0\rho_{c}}-1
\end{equation}

\noindent where $\rho_c$ is the critical density of the universe and $\Omega_0$ is the cosmological density parameter.

To describe the peculiar velocity pattern around galaxy clusters, Yahil (1985) introduced the following non-linear approximation

\begin{equation} \label{eq:Y}
\frac{v_{pec}(r)}{H_0r}=-\frac{1}{3}\Omega_0^{0.6}\frac{\Delta(<r)}{(1+\Delta(<r))^{0.25}}
\end{equation}

\noindent As a result, one can get the infall velocity profile, $v_{inf}(r)$, for a cluster as

\begin{equation} \label{eq:Y1}
\frac{v_{inf}(r)}{H_0r}=1-\frac{1}{3}\Omega_0^{0.6}\frac{\Delta(<r)}{(1+\Delta(<r))^{0.25}}
\end{equation}

The density contrast profile can be obtained by two different ways. The first way is to obtain 
$\Delta(<r)$ from the spatial number density profile of the cluster and the background number density of 
the universe. The second way is to determine the mass density profile of the cluster and the background mass 
density of the universe using $\rho_{bg}=\Omega_0\rho_c$ (van Haarlem et al. 1993).

The value of the density contrast at the turnaround radius can be derived using Eq. \ref{eq:Y1} 
where $v_{inf}=0$. Accordingly, $\Delta(r=r_t)\approx 4.62$ for $\Omega_0=1$.

Einstein and Straus (1945, 1946) determined the metric of space-time near a star embedded in an 
expanding universe without a cosmological constant within general relativity. They assumed a 
star of mass $M$ is surrounded by an empty spherical cavity, called Einstein-Straus vacuole, 
with a radius, $r_{ES}$, defined as

\begin{equation}
r_{ES}=\left(\frac{3M}{4\pi\rho_{bg}}\right)^{1/3}
\end{equation}

\noindent From the definition of the Einstein-Straus vacuole, it can be considered that 
$r_{ES}=r_t$ (for more details see \citealt{Bonnor87}, Plaga 2005). Therefore, one can calculate 
$r_{ES}$ theoretically by knowing the cluster mass. 

Praton \& Schneider (1994) presented a model to predict the infall velocity profile based on spherical accretion 
onto a central mass seed, $m_c$, in an otherwise uniform and expanding universe (see, e.g., Peebles 1980). 
Consider the material surrounding $m_c$ to be divided into nested spherical shells in which each shell is labeled 
by a parameter $\phi$, which is the the development angle of the Friedmann model (Regos \& Geller 1989). 
The larger $\phi$ is, the closer the shell lies to the cluster. For bound or collapsed shells, $\phi$ is positive 
and for unbound shells, $\phi$ is negative. The equations of motion for these shells is

      \begin{equation}
      r(\phi)=r_v \left[\frac{M(\phi)}{M_v}\right]^{1/3} \left[\frac{3\pi/2+1}{\phi-\sin(\phi)} \right]^{2/3} \left|1-\cos(\phi)\right| ,
      \end{equation}

      \begin{equation}
      \frac{dr}{dt}(\phi)=\pm \sqrt{3}\sigma_v \left[\frac{M(\phi)/M_{v}}{r(\phi)/r_v} \right]^{1/2}(1+cos(\phi))^{1/2},
      \end{equation}
      
\noindent where the negative sign is for $\phi<0$
      
      \begin{equation}
      \frac{M(\phi)}{M_v}=\frac{f(\Omega_0)^{2/3}+(3\pi/2+1)^{2/3}}{f(\Omega_0)^{2/3}\pm[\phi - \sin(\phi)]^{2/3}},
      \end{equation}

      \begin{equation}
      f(\Omega_0)\equiv 2\frac{\sqrt{1-\Omega_0}}{\Omega_0}-\cosh^{-1}\left(\frac{2-\Omega_0}{\Omega_0}\right),
      \end{equation}

\noindent where $\sigma_v$ is the line of sight velocity dispersion of the cluster at $r_v$.

Praton model can be applied if one have the turnaround radius and the line of sight velocity dispersion of a cluster. 
Although $r_v$ in the Praton model is somewhat arbitrary or ad hoc we apply the Praton model for $\Omega_o=1$ after 
getting $r_v$ and $\sigma_v$ observationally. This is because $r_v$ is easier to get from observation.

\begin{figure*} \centering
\includegraphics[width=20cm,height=22cm]{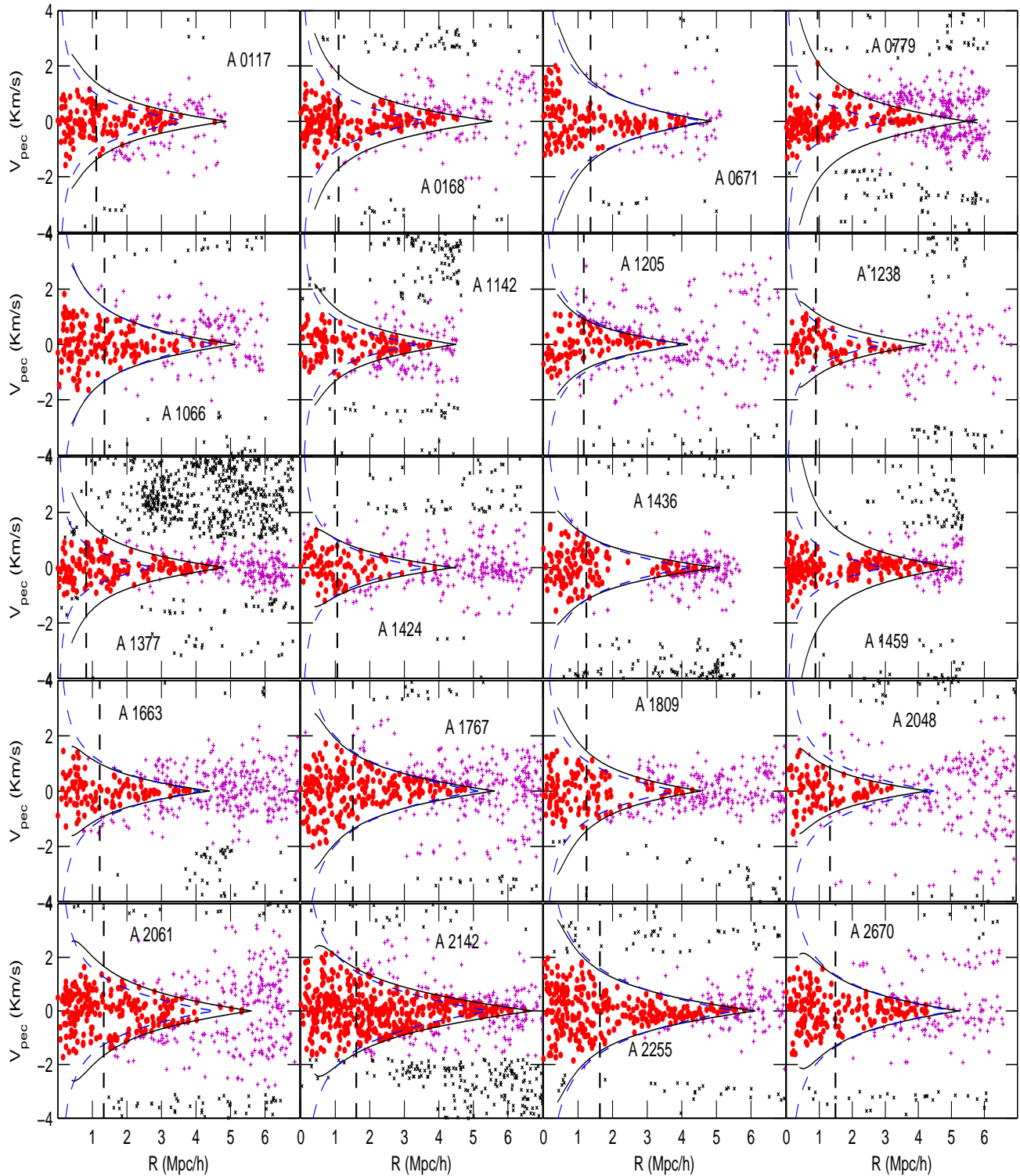} \vspace{-1cm}
\caption{Distribution of galaxies in redshift space. The filled points indicate 
the cluster members. The plus and cross symbols indicate the interlopers obtained by AKM and SIM, 
respectively. The solid curved lines are the location of the caustics using 
Yahil model. The dashed curved lines indicate the location of the caustics using Praton model. 
The vertical dashed line indicates $r_v$.}
\label{fig:RS} \end{figure*}

Figure \ref{fig:RS} shows the distribution of galaxies in the redshift space for each galaxy cluster. 
The filled points indicate the cluster members. The plus and the cross symbols refer to the interlopers 
obtained by AKM (first step) and interlopers obtained by the SIM (second step), respectively. 
The solid curved lines indicate the location of the caustics using Yahil model. The dashed curved lines 
indicate the location of the caustics obtained by Praton model for $r_v$. As mentioned before, we used 
Yahil approximation to determine the cluster members and the turnaround radius.

Yahil Approximation does not describe clusters' virialized regions because of the assumption of conservation of 
mass is not valid within the cluster core due to the crossing shells where there are mass transfer 
inside the virialized region. Notice that Yahil approximation and Praton model are nearly coincide for the 
clusters A0671, A1066, A1205, A1424, A1436, A1663, A1767, A2255, and A2670. Also, Yahil approximation is enclosed 
within the Praton model for only A2048, while Praton model is enclosed and/or matched with Yahil approximation 
for A0117, A1142, A1238, A1809, and A2061. Finally both models are unmatched at all for A0168, A0779, A1377, and A1459. 

For the two clusters A0779 and A1459, which are characterized by high central densities, Yahil approximation 
fails to describe their infall velocity profiles. In other words, the caustic boundaries of these two clusters 
are large in comparison with the distribution of their members within the virialized region. On the other hand, 
Praton model describes the infall velocity profiles for these two clusters with good acceptance. The two clusters 
A1436 and A1809 have nearly two empty regions between approximately $2-3.2$ and $\geq 2$, respectively. This may 
cause an overestimate in determination of the turnaround radii for these two clusters.
\section{Velocity Dispersion Profile} \label{sec:veldis}

\begin{figure*}
\includegraphics[width=18cm,height=15cm]{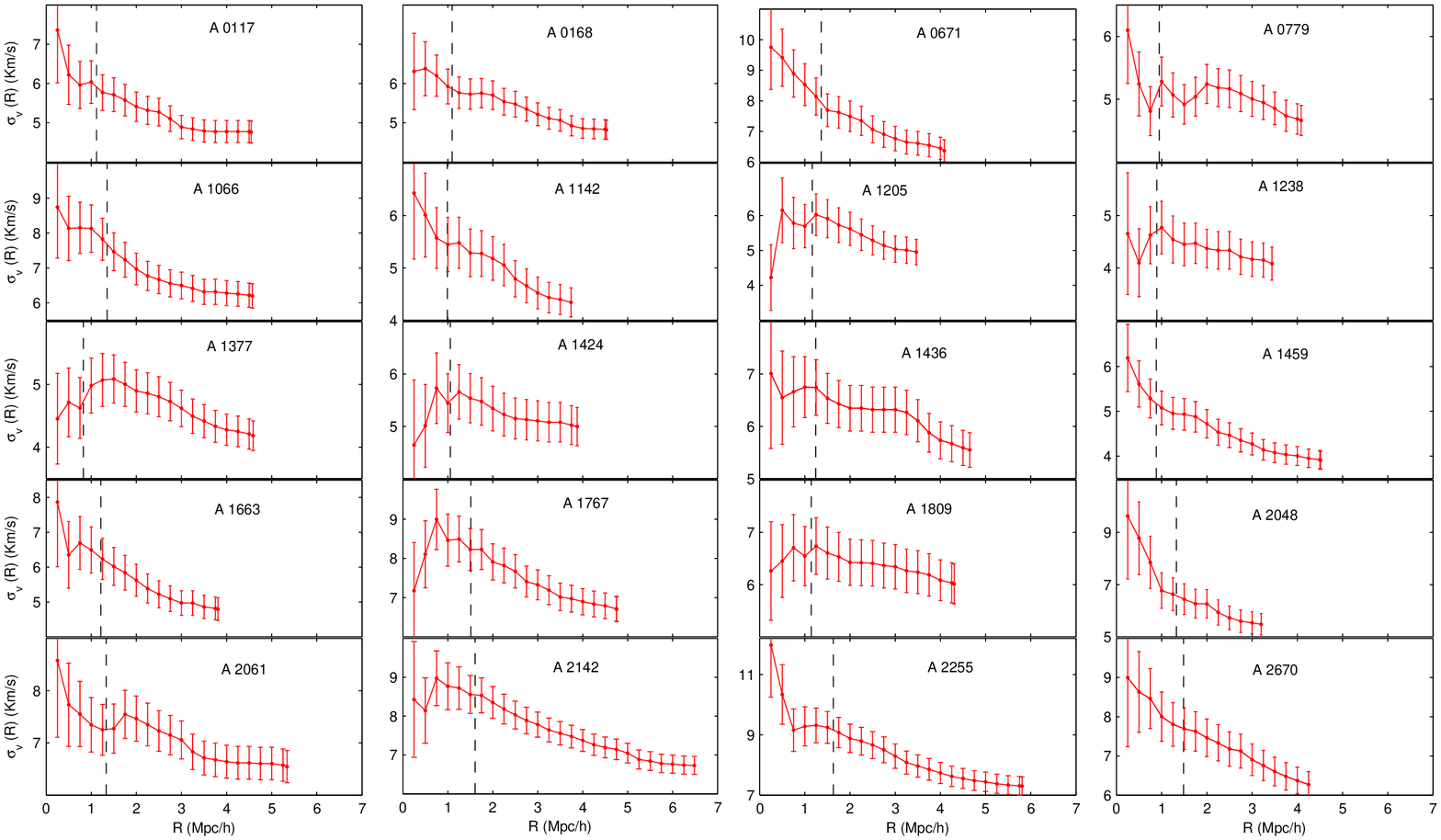}	
\caption{The Integrated velocity dispersion profile where the dispersion at a given radius 
is the average velocity dispersion within that radius. The vertical dashed line indicates 
the virial radius obtained by the corrected virial mass estimator.}
\label{fig:VD}\end{figure*}

The velocity dispersion profile, hereafter VDP, of a galaxy cluster is a measure of the 
cluster's dynamical state and is considered as the most important ingredient for the 
calculation of different mass estimators and mass profiles.

For each galaxy the factor $(1+z_{obs})$, with $z_{obs}$ is the observed redshift of the 
galaxy, is the product of a factor $(1+z_{cl})$, which is due to the cosmological redshift, 
and $(1+z_{gal})$, which is the Doppler term due to the velocity of the galaxy with respect 
to the cluster center, where $z_{cl}$ is the cluster redshift and $z_{gal}$ is galaxy redshift 
with respect to the cluster center (Harrison \& Noonan 1979, \citealt{Danese80}).

  \begin{equation} \label{eq:correction}
  (1+z_{obs})=(1+z_{cl})(1+z_{gal}) \end{equation}

\noindent Therefore, the peculiar velocities (or the velocity dispersions) of galaxies in the 
frame of the cluster, taking into account the correction for cosmological redshift, are computed as
  
  \begin{equation} \label{eq:vdcalculation}
  v_{pec}=\frac{c(z_{obs}-z_{cl})}{(1+z_{cl})} \end{equation}

den Hartog \& Katgert (1996) classified VDPs into three kinds; peaked, which deceases with 
the distance from the cluster center, flat, and inverted, which increases with the distance 
from the cluster center. The differences in VDPs are due to several factors like cluster 
member selection, choice of the cluster center, possible velocity anisotropies in galaxy 
orbits, presence of substructure and the presence of a population of spiral galaxies not in 
virial equilibrium with the cluster potential or the projection effect. 
Also, VDP may be influenced by the existence of other structures on larger scales such as a 
nearby cluster or the super-cluster, to which the cluster belongs, or filament and so on 
(see den Hartog \& Katgert 1996, Girardi et al. 1996, Fadda et al. 1996). 

Possible velocity anisotropies affect the shape of VDP, particularly within the central 
regions of the clusters. In order to avoid effects of possible anisotropies on the total 
value of velocity dispersion, Girardi et al. (1996) suggested studying the integral velocity 
dispersion profile (hereafter IVDP), where the dispersion at a given radius is evaluated by 
using all the galaxies within that radius. Although the presence of velocity anisotropies 
can strongly influence the value of integral velocity dispersion $\sigma$ computed for 
the central cluster region, it does not affect the value of the spatial (or projected) 
$\sigma$ computed for the whole cluster (The \& White 1986, Merritt 1988).

\begin{figure*}
\includegraphics[width=18cm,height=16cm]{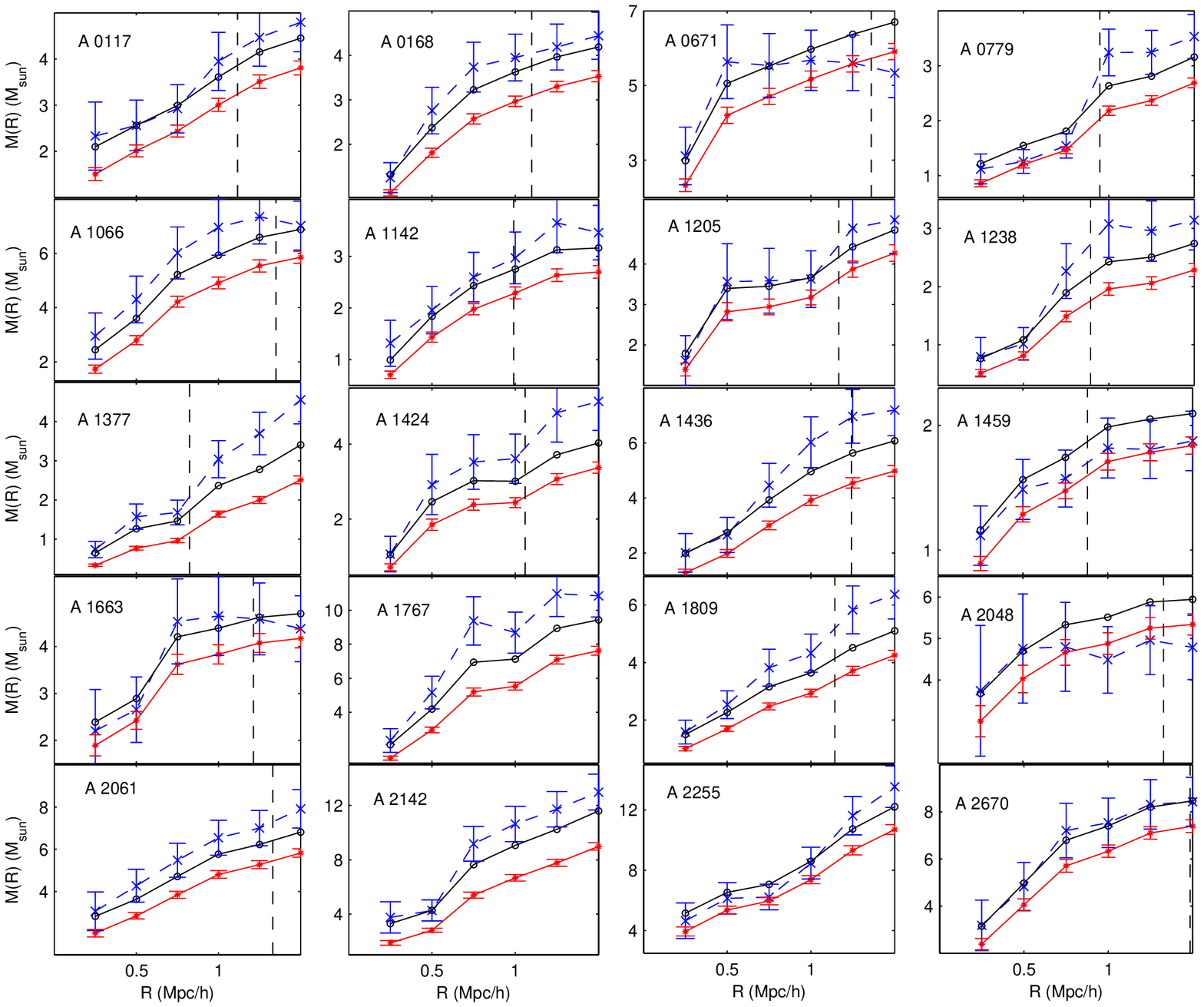}
\caption{The cluster mass profiles obtained using different mass estimators. The open circle and dotted lines 
are the virial mass profile before and after correction respectively. The cross line represents the mass profile 
obtained using the isotropic projected mass estimator. The vertical dashed line indicates the virial radius 
obtained by NFW mass profile.}
\label{fig:MP}\end{figure*}

IVDP for each galaxy cluster is presented in Figure \ref{fig:VD}. The vertical line indicates the virial 
radius. Generally, the trend of IVDP is decreasing with increasing distance from the cluster center for 
most of the clusters, but it sometimes exhibits irregularities in the virialized regions of some clusters. 
These irregularities are not due to the effect of substructures because the galaxy clusters have no evidence 
for substructures (see Figure \ref{fig:RS}). Also, although we neglect the effect of the presence of galaxy 
clusters in larger structures, IVDPs don't exhibit any odd trend in the outer regions.

The irregular trend of IVDP within the virialized region is shown for the 7 clusters, A1205, A1238, A1377, 
A1424, A1767, A1809 and A2142, which may be due to two main factors. First, the projection effect where the 
calculated velocity dispersion within the central region is affected by the outer members that appear in projection 
very near to the center in the projected distance but they are far in the redshift direction from the mean 
redshift of the cluster. In other words, the population of late type galaxies, which are usually 
exist in the outer region of clusters, may be found within the projected central region. 
To study this effect we have to identify the types of the galaxies in our sample which need long time to 
do, therefore we ignore that at this time. Second, the velocity anisotropy which is very poorly to identify. 
Because of determining IVDP not VDP, the calculation of mass, using the virial and projected mass 
estimators, outside the central regions will not be affected by this odd behavior.
\section{Masses and Mass Profiles of Galaxy Cluster} \label{sec:mass}
The Cluster mass profile can be determined using different methods. The following two methods are used throughout this work. 

\begin{table} 
\caption{Fitted parameters for the NFW mass profile.}
\label{tab:NFWfitting}
\begin{tabular}{|c|c|c|c|c|}
\hline
Name &      $r_s$  & $M_s$       & c              &R-Square \\                             
     &     (Mpc/h) &$(10^{13} M_{\odot})$&        &     \\\hline  
A0117&0.11 $\pm$ 0.04&4.1 $\pm$ 0.8&10 $\pm$ 4    &0.98 $\pm$ 0.01\\
A0168&0.15 $\pm$ 0.05&4.8 $\pm$ 1.0&7.4 $\pm$ 2.6 &0.98 $\pm$ 0.01\\
A0671&0.06 $\pm$ 0.02&5.2 $\pm$ 0.8&23 $\pm$ 7    &0.98 $\pm$ 0.01\\
A0779&0.12 $\pm$ 0.08&2.9 $\pm$ 1.1&8.0 $\pm$ 5.1 &0.95 $\pm$ 0.01\\
A1066&0.12 $\pm$ 0.03&7.1 $\pm$ 1.1&11 $\pm$ 3    &0.99 $\pm$ 0.01\\
A1142&0.11 $\pm$ 0.03&3.1 $\pm$ 0.4&9.1 $\pm$ 2.5 &0.99 $\pm$ 0.01\\
A1205&0.07 $\pm$ 0.04&3.8 $\pm$ 1.0&16 $\pm$ 8    &0.95 $\pm$ 0.02\\
A1238&0.18 $\pm$ 0.10&3.3 $\pm$ 1.3&5.0 $\pm$ 2.8 &0.93 $\pm$ 0.03\\
A1377&0.63 $\pm$ 0.22&9.4 $\pm$ 3.5&1.3 $\pm$ 0.5 &0.99 $\pm$ 0.01\\
A1424&0.15 $\pm$ 0.05&4.4 $\pm$ 1.0&7.0 $\pm$ 2.4 &0.98 $\pm$ 0.01\\
A1436&0.23 $\pm$ 0.05&8.4 $\pm$ 1.3&5.5 $\pm$ 1.2 &0.99 $\pm$ 0.01\\
A1459&0.03 $\pm$ 0.01&1.3 $\pm$ 0.1&33 $\pm$ 7    &0.99 $\pm$ 0.01\\
A1663&0.07 $\pm$ 0.05&4.0 $\pm$ 1.3&18 $\pm$ 12   &0.88 $\pm$ 0.04\\
A1767&0.31 $\pm$ 0.09&16.5 $\pm$ 4.1&4.9 $\pm$ 1.4&0.98 $\pm$ 0.01\\
A1809&0.19 $\pm$ 0.06&6.2 $\pm$ 1.5&6.0 $\pm$ 1.9 &0.98 $\pm$ 0.01\\
A2048&0.03 $\pm$ 0.02&3.9 $\pm$ 1.1&39 $\pm$ 26   &0.91 $\pm$ 0.03\\
A2061&0.11 $\pm$ 0.03&6.6 $\pm$ 0.9&12 $\pm$ 3    &0.99 $\pm$ 0.01\\
A2142&0.37 $\pm$ 0.11&2.2 $\pm$ 5.9&4.4 $\pm$ 1.3 &0.99 $\pm$ 0.01\\
A2255&0.08 $\pm$ 0.05&9.1 $\pm$ 2.9&21 $\pm$ 13   &0.92 $\pm$ 0.02\\
A2670&0.08 $\pm$ 0.02&7.4 $\pm$ 1.0&18 $\pm$ 4    &0.96 $\pm$ 0.01\\
\hline                       
\end{tabular} \end{table}

1. Virial Mass Estimation: Depending on the virial theorem, the masses of a galaxy cluster can be determined if it 
is assumed that they are bound, self-gravitating systems. The virial mass profile, $M_v(<r)$, can be evaluated by 
 
  \begin{equation} \label{eq:vir16}
  M_v(<r)=\frac{3\pi N \sum_{i}v_{pec, i} (<r)^2}{2G\sum_{i\neq j}\frac{1}{R_{ij}}} 
  \end{equation}

\noindent where $v_{pec}$ is the peculiar velocity of a galaxy (see Eq. \ref{eq:vdcalculation}) and $R_{ij}$ is the 
projected distance between two galaxies (see Limber \& Mathews 1960, \citealt{Aceves99a}, Rines et al. 2003). 
The uncertainty for the virial theorem is calculated using the limiting fractional uncertainty 
$\pi^{-1}(2\ln N)^{1/2}N^{-1/2}$ (\citealt{Bahcall81}). 

If velocity anisotropies exist or the assumption that mass follows light does not hold, the virial mass 
estimator may produce misleading results (The \& White 1986, Merritt 1988).If the system is extend beyond 
the virial radius, Eq. \ref{eq:vir16} overestimates the mass by external pressure from the matter outside 
the virialized region (The \& White 1986, Girardi et al. 1998, \citealt{Carlberg97}). Accordingly, it has to 
add an additional term to Eq. \ref{eq:vir16} called the surface pressure term, C(r). Thus the virial mass 
profile can be determined using the following expression

  \begin{equation} \label{eq:vir17}
  M_{vc}(<r)=M_v(<r)[1-C(r)], \end{equation}
  
  \begin{equation} \label{eq:vir18}
  C(r)=4\pi r^3 \frac{\rho(r)}{\int_0^r 4\pi r^2 \rho(\hat{r})d\hat{r}}\left[\frac{\sigma_v(r)}{\sigma(<r)}\right]^2   
  \end{equation}

\noindent where $\sigma(<r)$ is the integrated three-dimensional velocity dispersion 
within the radius r, $\sigma_v(r)$ is a projected velocity dispersion within that radius, 
and $\rho(r)$ is the density distribution (Koranyi \& Geller 2000; Tustin et al. 2001).  

For NFW density profile and for isotropic orbits of galaxies inside the cluster

\begin{equation}\label{eqn:vir_25}
C(r)=\frac{(r/r_s)^2}{(1+r/r_s)^2}\left[\ln(1+\frac{r}{r_s})-\frac{r/r_s}{1+r/r_s}\right]^{-1}\left[\frac{\sigma_v(r)}{\sigma(<r)}\right]^2
\end{equation}

2. Projected Mass Estimation: An alternative method, projected mass method, is discussed by many 
authors e.g. Page (1952), Wolf \& Bahcall (1972), \citealt{Bahcall81}, Heisler \& Tremaine (1985) 
and \citealt{Aceves99b}. It depends on the distribution of the orbits of galaxies around the cluster center. 
The isotropic projected mass can be written as

  \begin{equation} \label{eq:pro7}
  M_{PI}(<R)=\frac{32}{\pi GN}\sum_{i}v_{pec,i}^2R_i , 
  \end{equation} 

\noindent If the orbits are purely radial or purely circular, the factor 32 becomes 64 or 16, 
respectively (Rines et al. 2003).

Using N-body simulations Perea et al. 1990 conclude that, among the two mass estimators the virial 
mass estimator is the best method and it is unaffected by the presence of substructure or anisotropies. 
However, it is affected by the presence of interlopers and the existence of mass distribution. 
The projected mass estimator is largely affected by the presence of anisotropies, the existence 
of substructure or the presence of interlopers. They also demonstrate that any factor that was not 
taken into account would give an overestimation of the mass of the system by factors between 2 to 4.

\begin{table*}
\caption{Cluster parameters at the virial radius.}
\label{tab:rvprop}
\begin{tabular}{|c|cccc|ccccc|cc|}
\hline
Name&\multicolumn{4}{c||}{$r_v$ (Mpc/h)}&$No$&$\sigma$ (Km/s)&\multicolumn{5}{c||}{Mass $(\leq r_v) (10^{14}M_{\odot})$}\\
\cline{2-5} \cline{8-12}
&&AG2007&R2006&Pop2007&$(\leq r_v)$&$(\leq r_v)$& $M_p$ & $M_v$ & $M_{vc}$ &$M_{Pr}$ &$M_{NFW}$\\\hline 
A0117	&1.12	&0.89	&-    	&1.05	&77	&571 $\pm$ 65 &4.19 $\pm$ 0.60	&3.86	&3.24 $\pm$ 0.14	&3.26	&3.23 \\
A0168	&1.10	&0.89	&0.95 	&1.05	&102&582 $\pm$ 58 &4.05 $\pm$ 0.51	&3.76	&3.10 $\pm$ 0.11	&3.84	&3.08 \\
A0671	&1.37	&1.07	&1.12 	&-   	&99	&786 $\pm$ 79 &5.48 $\pm$ 0.70	&6.53	&5.69 $\pm$ 0.21	&6.14	&5.93 \\
A0779	&0.95	&0.59	&0.92 	&-   	&86	&468 $\pm$ 50 &2.88 $\pm$ 0.39	&2.46	&2.03 $\pm$ 0.08	&2.46	&1.97 \\
A1066	&1.35	&1.28	&1.26 	&1.26	&91	&764 $\pm$ 80 &7.22 $\pm$ 0.96	&6.71	&5.67 $\pm$ 0.22	&6.16	&5.72 \\
A1142	&0.99	&1.00	&0.98 	&-   	&55	&550 $\pm$ 74 &2.95 $\pm$ 0.50	&2.74	&2.28 $\pm$ 0.12	&2.79	&2.26 \\
A1205	&1.17	&1.53	&0.96 	&1.54	&50	&598 $\pm$ 85 &4.50 $\pm$ 0.80	&4.19	&3.66 $\pm$ 0.19	&3.06	&3.76 \\
A1238	&0.89	&0.98	&-    	&-   	&42	&471 $\pm$ 73 &2.73 $\pm$ 0.53	&2.20	&1.76 $\pm$ 0.10	&1.93	&1.66 \\
A1377	&0.82	&1.10	&0.83 	&-   	&52	&490 $\pm$ 68 &2.08 $\pm$ 0.36	&1.73	&1.16 $\pm$ 0.06	&2.37	&1.30 \\
A1424	&1.06	&1.10	&1.06 	&1.19	&53	&568 $\pm$ 78 &3.90 $\pm$ 0.68	&3.18	&2.58 $\pm$ 0.13	&2.93	&2.76 \\
A1436	&1.24	&1.13	&0.72 	&-   	&81	&674 $\pm$ 75 &6.94 $\pm$ 0.97	&5.62	&4.53 $\pm$ 0.19	&4.67	&4.46 \\
A1459	&0.90	&0.89	&-    	&-   	&86	&516 $\pm$ 56 &1.72 $\pm$ 0.23	&1.89	&1.68 $\pm$ 0.07	&2.74	&1.67 \\
A1663	&1.21	&1.16	&1.20 	&1.33	&56	&629 $\pm$ 84 &4.60 $\pm$ 0.78	&4.59	&4.04 $\pm$ 0.20	&3.56	&4.15 \\
A1767	&1.51	&1.53	&1.37 	&1.54	&122&825 $\pm$ 75 &10.9 $\pm$ 1.25  &9.47 &7.65 $\pm$ 0.26  &7.82 &8.02 \\
A1809	&1.14	&1.26	&0.83 	&1.19	&75	&680 $\pm$ 79 &5.19 $\pm$ 0.76	&4.14	&3.38 $\pm$ 0.15	&4.32	&3.47 \\
A2048	&1.33	&-   	&-    	&-   	&58	&658 $\pm$ 86 &4.91 $\pm$ 0.81	&5.90	&5.28 $\pm$ 0.26	&3.82	&5.45 \\
A2061	&1.33	&1.07	&1.31 	&-   	&116&725 $\pm$ 67 &7.29 $\pm$ 0.86	&6.41	&5.45 $\pm$ 0.19  &6.05 &5.48 \\
A2142	&1.61	&-   	&1.13 	&-   	&162&845 $\pm$ 66 &13.94 $\pm$ 1.38	&12.36&9.65 $\pm$ 0.28	&9.83	&9.69 \\
A2255	&1.63	&1.40	&1.48 	&1.82	&158&917 $\pm$ 73 &13.54 $\pm$ 1.36	&12.41&10.9 $\pm$ 0.33	&10.54&10.1 \\
A2670	&1.49	&1.10	&0.93 	&1.40	&103&778 $\pm$ 77 &8.42 $\pm$ 1.05	&8.45	&7.38 $\pm$ 0.27	&6.50	&7.63 \\
\hline 
\end{tabular} \end{table*} 

Under the assumption that the mass follows light, the two mass estimators (virial mass and isotopic projected mass) 
are used to derive the different mass profiles for each galaxy cluster. In Figure \ref{fig:MP} we plotted 
the uncorrected virial mass profile (open circle line), the corrected virial mass profile (dotted line), and
the isotropic projected mass profile (cross dashed line). On average, the surface pressure term reduces the virial 
mass estimation by about $14\%$. The isotropic projected mass profiles are higher than the corrected virial mass 
profiles within the virialized region. 

Depending on the virial mass profile, just within $1.5 \mbox{ Mpc} \mbox{ h}^{-1}$ to avoid the systematic error 
caused by the determination of the mass in the outer regions using virial theorem, we fit it with NFW profile 
(see Table \ref{tab:NFWfitting}). Cols. 1-4 are the scale radius, $r_s$, the mass within $r_s$, $M_s$, the 
concentration parameter, $c=r_v/r_s$ and the adjusted R-Square, respectively, (see NFW, Koranyi \& Geller 2000). 
The scale radius, $r_s$, of the studied sample has mean value $0.16 \mbox{ Mpc}\mbox{ h}^{-1}$ and ranges 
from $0.03 \mbox{ Mpc}\mbox{ h}^{-1}$ to $0.63 \mbox{ Mpc}\mbox{ h}^{-1}$. While the mass within $r_s$ has mean 
value $6.7 \times10^{13} M_{\odot}$ and ranges from $1.25 \times10^{13} M_{\odot}$ to $16.5 \times10^{13} M_{\odot}$. 
The concentration parameter, $c$, has mean value $12.98$ and range from $1.3$ to $39.17$, in good agreement with 
R2006 and with the predictions of numerical simulations (Navarro et al.1997, \citealt{Bullock01}).
\section{Clusters Parameters and their correlations} \label{sec:parameters}

In this section we introduce the clusters physical parameters and compare them with the results 
in the literature. We also investigate the correlation between the different parameter.

In Table \ref{tab:rvprop} we listed the clusters parameters at the virial radius. Col. 2 is $r_v$ obtained by NFW mass profile.
Cols. 3-5 are $r_v$ determined by AG2007 and R2006, and Popesso et al. 2007, hereafter Pop2007,  respectively. Cols. 6-12 
are the number of galaxies, $No(\leq r_v)$, the velocity dispersion, $\sigma(\leq r_v)$, the isotropic projected mass, 
$M_p(\leq r_v)$, the uncorrected virial mass, $Mv(\leq r_v)$, the corrected virial mass, $M_{vc}(\leq r_v)$, the mass calculated 
from praton model, $M_{Pr}(\leq r_v)$, and NFW mass, $M_{NFW}(\leq r_v)$, within $r_v$, respectively. AG2007 determine the 
cluster members using a different method and calculate $r_v$ using the relation $\sqrt{3} \sigma_c/10 H(z_c)$, where 
$\sigma_c$ is the cluster dispersion velocity and $z_c$ is the cluster redshift. R2006 calculate $r_v$ using the caustic mass 
profile (see Diaferio 1999) and Pop2007 calculate it using the virial mass profile. The ratios between our $r_v$ and AG2007, 
R2006 and Pop2007 are, respectively, $1.09\pm 0.21$, $1.12\pm 0.23$ and $0.96\pm 0.10$. There is good agreement between these 
studies for the most clusters. However, A0779, A1205, A1377 and A2670 have $r_v$ far from AG2007. Also A1436, A1809, A1242, 
A2670 have $r_v$ far from R2006. Moreover, A1205 have far $r_v$ from Pop2007. The differences between $r_v$ for these four 
studies are clearly due to the method used to get cluster members and the method used to get cluster mass profile.

The ratios between $M_{vc}$ and $M_p$, $M_{Pr}$ and $M_{NFW}$  are averaged and give $0.78\pm 0.14$, $0.93\pm 0.20$ and $0.99\pm 0.04$, respectively. 
$M_{vc}$ at $r_v$ has good agreement with $M_{Pr}$ and $M_{NFW}$, while $M_P$ is larger than the others. To compare the determined 
corrected virial mass with R2006 \& Pop2007, we get the mass from the corrected viral mass profile at each of $r_v$ of these two 
studies. For R2006 we find that the ratio of ours and the mass calculated from the caustic method at $r_v$ of this study is 
$1.58\pm0.77$ and for the viral mass estimator is $1.52\pm0.50$, which show that our results are larger than R2006 by about $58\%$ 
and $50\%$ for caustic and viral mass estimators, respectively. Also, the ratio of ours to Pop2007 at $r_v$ of this study is 
0.$65\pm0.16$, which indicates that our results give lower masses than Pop2007 by about $35\%$.

\begin{table} \centering 
\caption{Cluster parameters at the turnaround radius.} \label{tab:rtprop}
\begin{tabular}{ccccc}\hline
Name&$r_t$&$r_{tP}$&$r_{tN}$&$r_{ES}$ \\       
&\multicolumn{4}{c}{$\mbox{(Mpc/h)}$}  \\\hline
A0117  &4.88	&3.80	&4.86	&8.06	\\
A0168  &5.54	&3.74	&4.93	&8.28	\\
A0671  &4.86	&4.65	&5.63	&9.24 \\
A0779  &5.48	&3.22	&4.21	&7.20	\\
A1066  &5.14	&4.59	&5.84	&9.56	\\
A1142  &4.50	&3.37	&4.36	&7.25	\\
A1205  &4.18	&4.00	&4.93	&8.07	\\
A1238  &4.24	&3.04	&4.18	&6.94	\\
A1377  &4.82	&2.80	&4.90	&8.09	\\
A1424  &4.43	&3.60	&4.77	&7.84	\\
A1436  &5.01	&4.23	&5.74	&9.35	\\
A1459  &5.03	&3.05	&3.64	&6.20	\\
A1663  &4.39	&4.13	&5.07	&8.30	\\
A1767  &5.61	&5.14	&7.02	&11.37\\
A1809  &4.56	&3.89	&5.23	&8.52	\\
A2048  &4.31	&4.52	&5.31	&8.73	\\
A2061  &5.61	&4.53	&5.74	&9.50	\\
A2142  &6.72	&5.48	&7.64	&12.45\\
A2255  &6.12	&5.55	&6.78	&11.13\\
A2670  &5.24	&5.06	&6.21	&10.16\\
\hline
\end{tabular} \end{table}

In Table \ref{tab:rtprop} we listed, the parameters calculated at the turnaround radius, $r_t$. Col. 2 gives $r_t$ at 
which Yahil infall velocity goes to zero, col. 3 gives the turnaround radius $r_{tP}$ obtained from Praton model for $\Omega_0=1$ 
where $r_{tP}\approx 3.4 r_v$, cols. 4 is the turnaround radius, $r_{tN}$, obtained from NFW mass profile and equation 8 of 
Regos \& Geller (1989) for $\Omega_0=1$, i.e. at density  $3.55 \rho_c$ and col. 5 gives the calculated Einstein-Straus radius, 
$r_{ES}$, using the NFW mass, $M_{NFW}(\leq r_t)$ for $\rho_m=\rho_c$ where $\Omega_o=1$. The ratios of $r_t$ to $r_{Pr}$ and 
$r_{tN}$ are, respectively, $1.25\pm0.23$ and $0.96\pm0.16$ which show good agreement between them. The ratio 
$r_t/r_{ES} = 0.58\pm 0.08$ which means that $r_t$ is lower than $r_{ES}$ by about $42\%$. This is due to the choice of 
$\rho_m$ to be equal to $\rho_c$ for $\Omega_o=1$. Instead, $\rho_m$ should be greater than $\rho_c$ in order to the 
structure formation start to form.
\section{Conclusion} \label {sec:conc}
We used DR7 of SDSS to construct a sample of 42382 galaxies with redshifts in the region of 20 galaxy clusters. 
We distinguished between interlopers and cluster members by using two iterative steps, the adaptive kernel method 
and SIM, respectively. Consequently, we obtained 3396 galaxy members belonging to the studied cluster sample. 
We presented the two-dimensional optical maps of the studied sample using the adaptive kernel method to determine 
their centers. The cumulative surface number density profile is fitted well with the generalized King model. The core 
radius varies from 0.18 Mpc $\mbox{h}^{-1}$ (A1459) to 0.47 Mpc $\mbox{h}^{-1}$ (A2670) and has mean value of 0.295 Mpc 
$\mbox{h}^{-1}$. The velocity distribution for each cluster appears as a well-isolated peak with Gaussian distribution 
which means that the studied clusters have no substructures that influence the different dynamical properties of 
the galaxy clusters. 

The infall velocity profile of each cluster was determined using two different models: Yahil approximation and Praton 
model. We confirm that Yahil approximation can be applied only in the outskirt of the cluster far from the central 
virialized region, because the assumption of mass conservation is not valid. On the other hand, Praton model can be 
applied within the virialized region. The infall velocity determined by Praton model is matched with that determined 
by Yahil approximation in the outskirts of most studied clusters in the sample but they are unmatched for the clusters 
characterized by high central density. Yahil approximation is not valid for those clusters, while Praton model can 
describe the infall pattern for them with good approximation.

The integrated velocity dispersion profiles show that there are some irregularities in the profiles within 
the cluster's virial radius, while all profiles exhibit a flattened out behavior outside the virial radius. 
The two main factors caused this behavior are the projection effect and velocities anisotropies.
Under the assumption that the mass follows galaxy distribution, we determine the mass and mass profile by 
two independent mass estimators; projected mass and virial mass methods. The virial mass profile 
is corrected by applying the surface pressure term which reduces the virial mass by about $14\%$. The projected 
mass profile is larger than the corrected virial mass profile for nearly all clusters by about $28\%$. The virial 
mass agree with NFW mass and Praton mass at $r_v$. The virial mass profile within 1.5 Mpc $\mbox{h}^{-1}$ is fitted 
with NFW mass profile. The concentration parameter ranges from $1.3$ to $39.17$, and has mean value $12.98$ in good 
agreement with R2006 and with the predilections of numerical simulations (Navarro et al.1997, \citealt{Bullock01}).

Our great sincere thanks are for Dr. Elizabeth Praton (Dept. of Physics \& Astronomy, Franklin \& Marshall College, 
Lancaster, USA) for provide us with some Mathematica notebooks and for her help to explain some important points.

\end{document}